\documentstyle[11pt,newpasp,twoside,epsf]{article}
\def\edcomment#1{\iffalse\marginpar{\raggedright\sl#1\/}\else\relax\fi}
\reversemarginpar

\begin{document}
\title{Star Formation in Satellite Galaxies}
 \author{Jos\'e G. Funes, S.J.}
\affil{Vatican Observatory, Univ. of Arizona, Tucson, AZ 85721, USA}
\author{Carlos M. Guti\'errez}
\affil{Instituto de Astrof\'\i sica de Canarias, E-38200 La Laguna, Tenerife,
Spain}
\author{Francisco Prada}
\affil{Instituto de Astrof\'\i sica de Canarias \& Isaac Newton Group of
Telescopes, Spain}
\author{Marco Azzaro}
\affil{Isaac Newton Group of Telescopes, Santa Cruz de La Palma, Spain}
\author{Marcelo B. Ribeiro}
\affil{Instituto de F\'\i sica, Universidade Federal do Rio de Janeiro, Brazil}

\begin{abstract}
The study of satellite galaxies can provide information on the merging and
aggregation processes which, according to the hierarchical clustering
models, form the larger spiral galaxies we observe. With the aim of
testing hierarchical models of galaxy formation, we have conducted an
observational program which comprises H$\alpha$ imaging for both the parent and
the satellite
galaxies, taken from the compilation by Zaritsky et al. (1997) that
contains 115 galaxies orbiting 69 primary isolated spiral galaxies. We have
observed a subsample of 37 spiral and
irregular galaxies taken from the compilation mentioned above. The aim of this
study is to determine star formation
properties of the sample galaxies. In this work we present the preliminary
results of this program that we have carried out with the 1.8-m
Vatican Telescope (VATT).
\end{abstract}

\section{Introduction}
The hierarchical galaxy formation scenario implies that galaxies are formed by
merging and accretion of previously collapsed systems. This framework has been
successful in predicting the clustering of galaxies as a function of redshift
(Silk 2000); however semianalytic models (Kauffmann, White, \& Guiderdoni
1993) and numerical simulation (Klypin et al.1999; Moore et al.1999) of
hierarchical galaxy formation predict satellite number counts an order of
magnitude larger than the observed counts in the Local Group. The problem of
the missing number of observed satellites may be related to processes that
limit the star formation. For example, the gas accretion by low-mass halos can
be squelched by a strong photoionizing background (Somerville 2002).  

The formation and evolution of satellite galaxies poses questions that are
still open:

$\bullet$ What are the relationships between the morphologies of the parent
galaxies and their satellite populations?

$\bullet$ How do the kinematic, dynamical, and chemical properties of the
satellites compare to those of the parent galaxies?

$\bullet$ How do the properties of the satellites vary with projected distance
from the primary galaxy?

$\bullet$ What are the star formation histories of the satellite galaxies? Are
they related to the star formation history of the primary galaxies? 

$\bullet$ What is the role of interactions among satellites or between
satellites and their parent galaxies in the evolution of the system? 

In this work we present a project aimed at addressing those questions related
to the star formation in satellites of isolated spiral galaxies. 

\section{Observational Program}
We have conducted an
observational program which comprises broadband photometry in the optical
and in the H$\alpha$ narrow-band for satellites of
external giant spiral galaxies (Guti\'errez, Azzaro, \& Prada 2002). In this
work we present the preliminary results of an H$\alpha$ imaging program aimed
at studying the star formation properties of satellites. 

\subsection{The Sample} 
The galaxies have been selected from the catalog by Zaritsky et al (1997). This
catalog is the most complete compilation of satellite galaxies in the
literature and contains 115 galaxies orbiting 69 primary isolated spiral
galaxies. The limiting magnitude for this catalog is $M_B$ $\sim$ 15.5. The
catalog includes satellites:  

$\bullet$ that are 2.2 magnitude fainter than the parent galaxy

$\bullet$ with a projected distances $\Delta$D = 500 kpc from their primaries 

$\bullet$ with a difference in recessional velocity between the parents and the
satellites of $\Delta v$ = 500 $kms^{-1}$  
	
Guti\'errez et al (2002) have observed a subsample of 60 satellites taken from
the catalog mentioned above. This sample includes most of the
objects situated at decl. $\geq$ - 22. 
The satellites span a large range of morphologies and surface brightness
profiles. Hubble types vary from elliptical to irregular with 35 objects were
classified as spirals. For each galaxy the effective radius, the scale radius
and the B/D ratio are tabulated. 
The broadband photometry will be published in a forthcoming paper (Guti\'errez
et al., in preparation). 

\subsection{H$\alpha$ Imaging}
We have obtained H$\alpha$+[N~{\scriptsize II}] emission-line images for a
subsample of 37 spiral and irregular
satellite galaxies. The narrow-band images of the sample were acquired during
three runs in December 2001, May 2002, and December 2002 with the 1.8-m
Vatican Advanced Technology Telescope (VATT) at the Mt. Graham International
Observatory. A back illuminated 2048 x 2048 Loral CCD was used as the detector
at the aplanatic Gregorian focus, f/9. It yielded a field of view was
6$.\hspace{-1.5pt}'$4 x 6$.\hspace{-1.5pt}'$4 with an image scale of 0.4
pixel$^{-1}$ after 2 x 2 pixel binning. The seeing varied between
0$.\hspace{-2pt}''$9 - 1$.\hspace{-2pt}''$8 with a typical value of
1.$\hspace{-2pt}''$2.

For each galaxy we have obtained 3 x 1800 second narrow-band images using a set
of interference filters with 70 \AA\ widths that  isolate the spectral region
characterized by the redshifted H$\alpha$ and [N~{\scriptsize
II}]$\,\lambda6548, 6583$ \AA\ emission lines. To
subtract the stellar continuum we subtracted a 1200-second R-band image,
suitably scaled, from the narrow-band image. The mean scale factor for the
continuum image was estimated by comparing the intensity of the field stars in
the two bandpasses.
The data from the two first runs have been reduced. The images will be
flux-calibrated using the spectrophometric standard stars observed during the
different runs. 

\begin{figure}
\plottwo{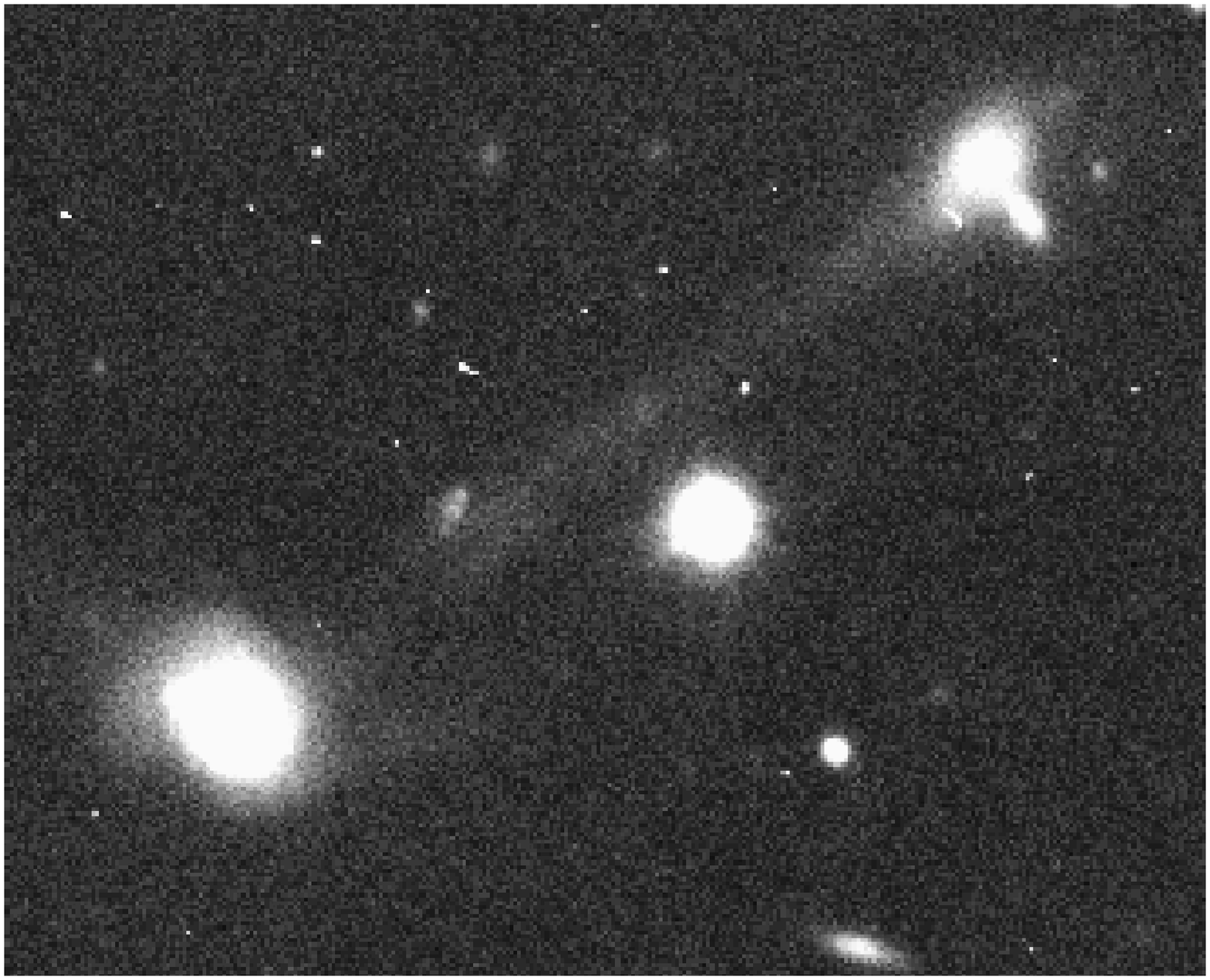}{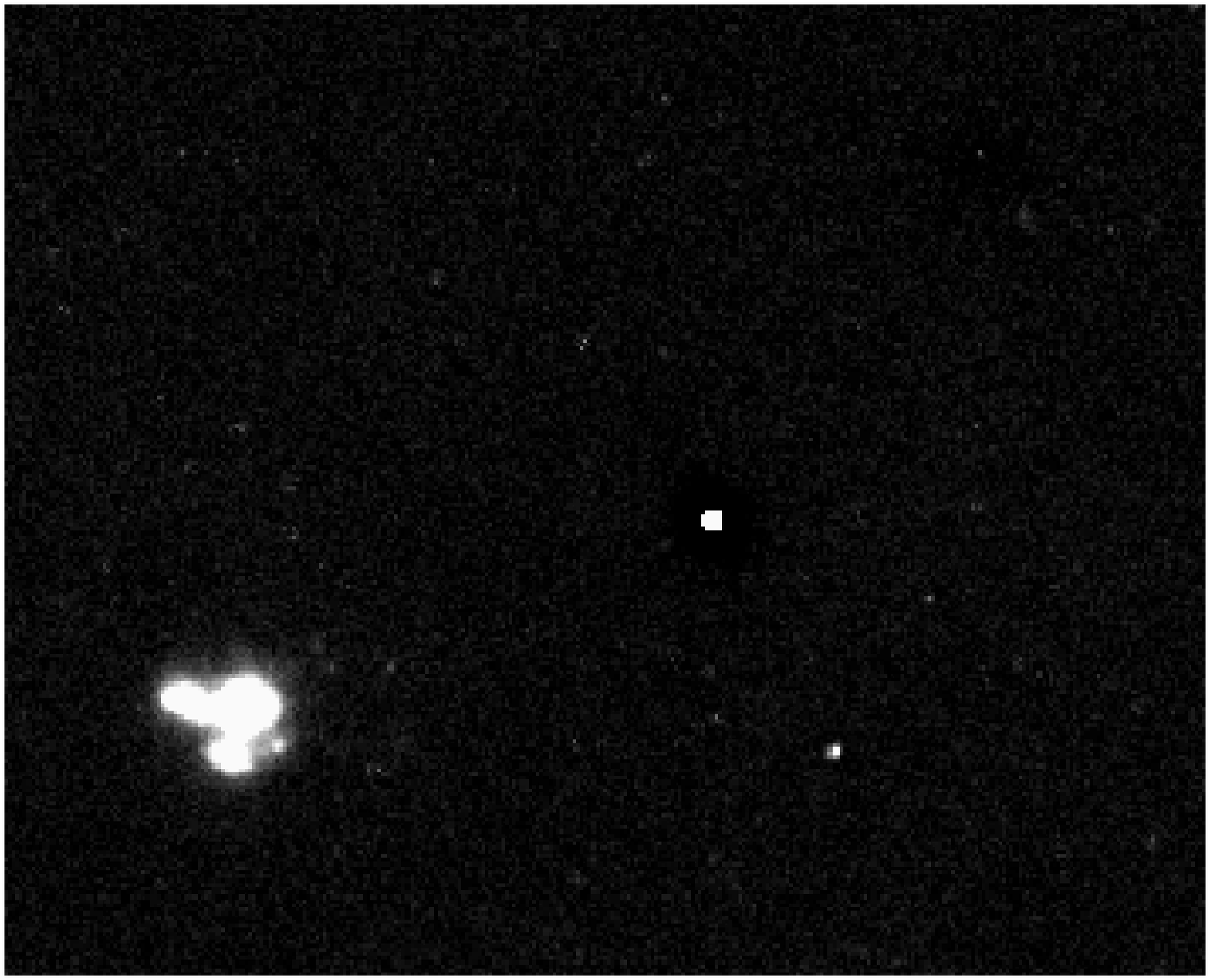}
\caption{{\it Left panel:} $R$-band image. NGC 2718a is on left-lower corner
and NGC 2718b is on the opposite corner. {\it Right panel:} continuum
subtracted H$\alpha$+[N~{\scriptsize II}] image. In both panels the
field of view is 2$.\hspace{-1.5pt}'$0 x 1$.\hspace{-1.5pt}'$6.North is up and
east left.}
\end{figure}

\section{Interaction Among Satellites}
Gutierrez et al. (2002) pointed out two cases of interaction in pairs of
satellites. In Figures 1 and 2 we show these cases.

{\bf NGC 2718a and NGC 2718b.} In the $R$-band image (Figure 1, right panel), a
bridge connecting the satellites is evident.  In the H$\alpha$+[N~{\scriptsize
II}] image (Figure 1, left panel), line emission is only observed in NGC
2718a, with no evidence for emission along the bridge. 

{\bf NGC 4541b and NGC 4541e.} Two tidal tails emerging from NGC 4541b are seen
in the $R$-band image (Figure 2, right panel). One tail is pointing to NGC
4541e and the other one
is in the opposite direction. In the H$\alpha$+[N~{\scriptsize II}] image
(Figure 2, left panel) there is no line emission seen in NGC 4541b.
 
In addition, our preliminary results indicate that interacting satellite
galaxies exhibit a
higher level of star formation than satellites with no signs of interaction.
Our work extends to satellite galaxies the idea - established previously for
luminous galaxies (Kennicutt et al. 1987)- that interactions trigger star
formation.  On the other side it would be interesting to search for starburst
activity in the parent galaxies produced by minor mergers as suggested by
numerical simulations (Mihos \& Hernquist 1994).

\begin{figure}
\plottwo{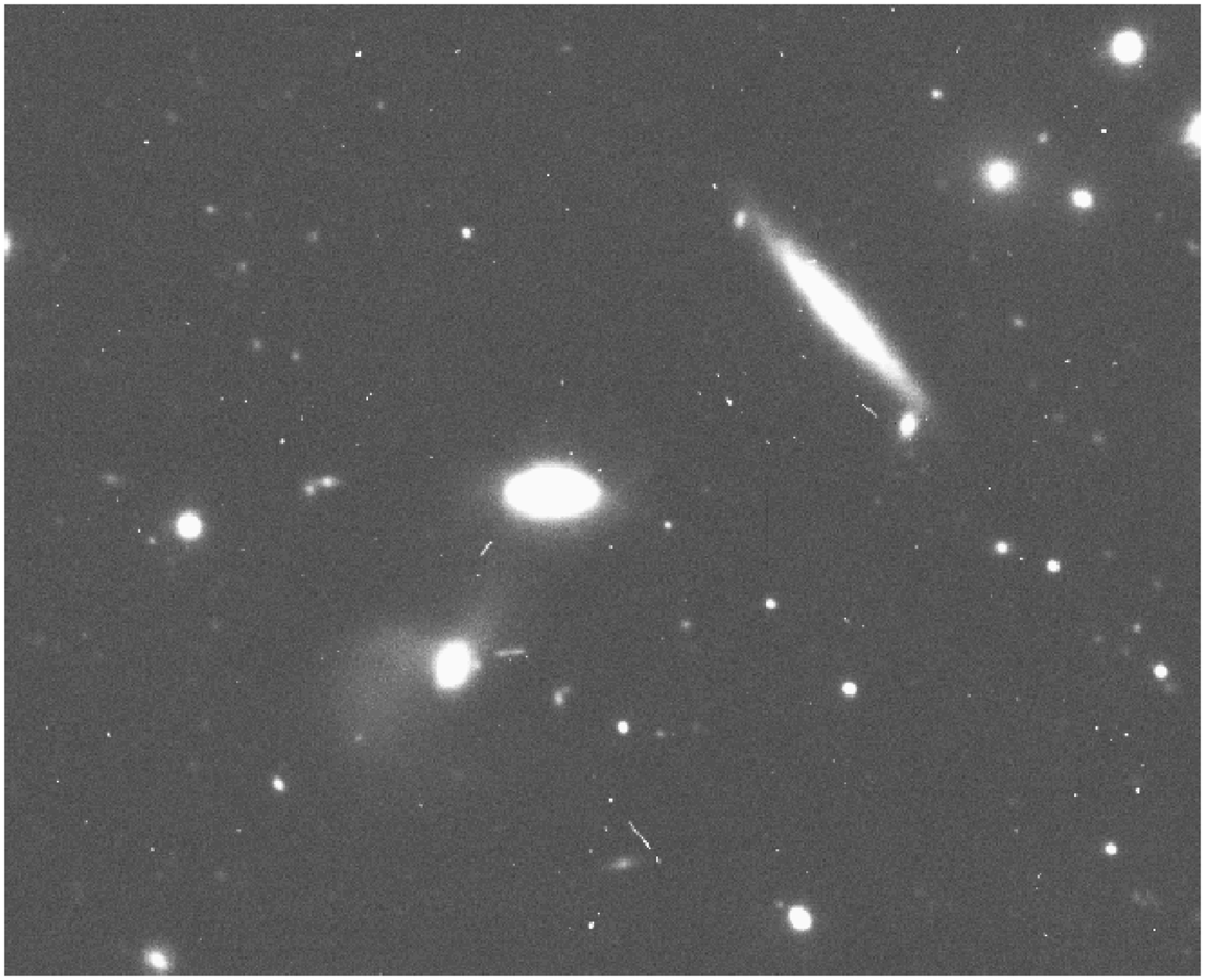}{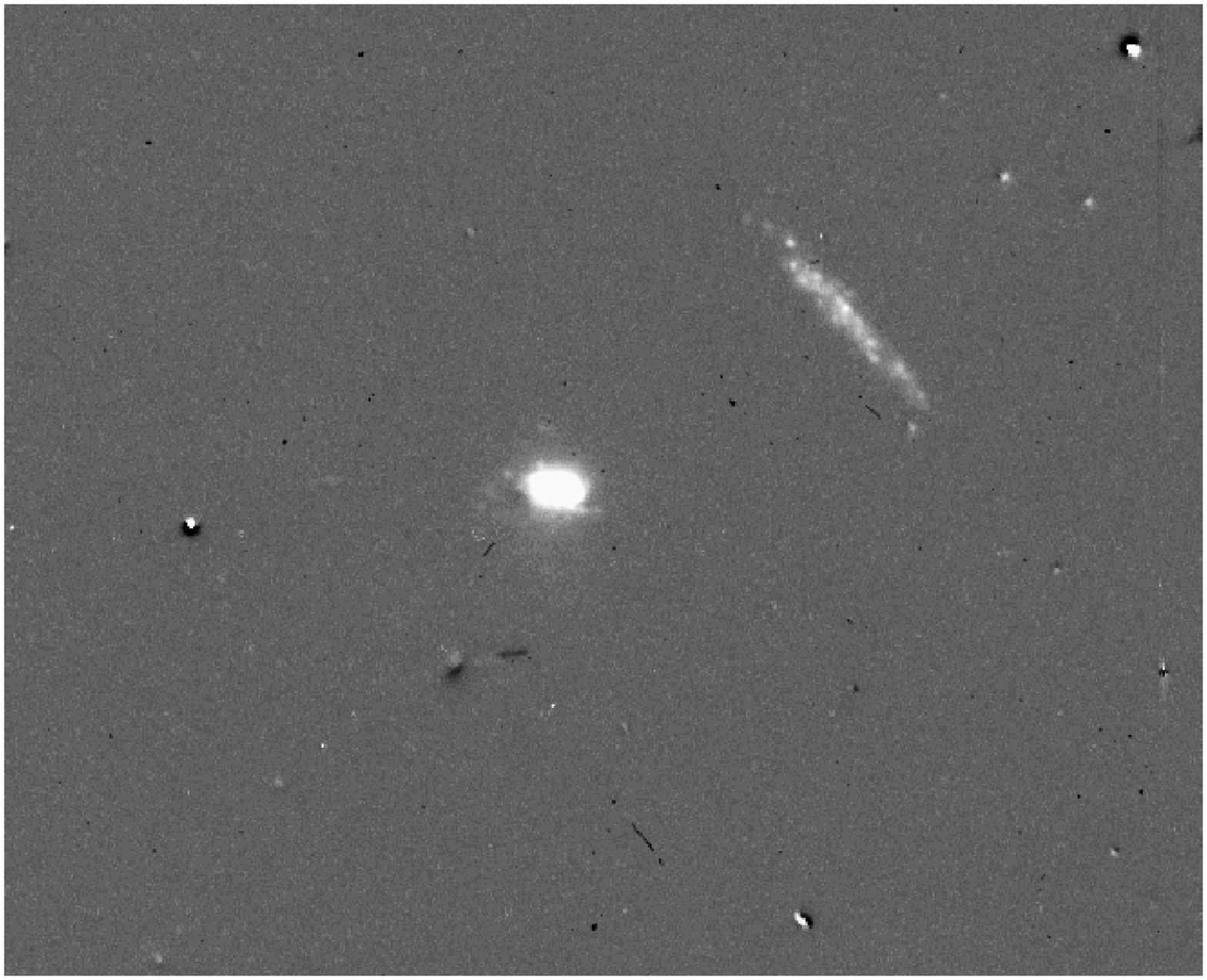}
\caption{{\it Left panel:} $R$-band image. NGC 4541a is the spiral on the
right-upper corner. NGC 4541e is in the middle and NGC 4541b is on the
left-lower corner. {\it Right panel:} continuum subtracted 
H$\alpha$+[N~{\scriptsize II}] image. In both
panels the field of view is 4$.\hspace{-1.5pt}'$0 x 3$.\hspace{-1.5pt}'$2.
North is up and east left.}
\end{figure}

\end{document}